\documentclass[a4paper,twocolumn,superscriptaddress,11pt]{quantumarticle}
\usepackage{epsfig,epstopdf}
\usepackage{graphics}
\usepackage[numbers,compress]{natbib}
\usepackage[english]{babel}
\usepackage[T1]{fontenc}
\usepackage[utf8]{inputenc}
\usepackage{braket,textcomp}
\usepackage[usenames]{color}
\usepackage[dvipsnames]{xcolor}
\usepackage[breaklinks, colorlinks=true, urlcolor=blue, 
anchorcolor=blue, citecolor=blue, filecolor=blue, linkcolor=blue, 
menucolor=blue, linktocpage=true, pdfproducer=medialab, 
pdfa=true]{hyperref}
\usepackage{amssymb,amsmath,bbm,mathcomp}
\usepackage{amsfonts,amsthm,subfigure}
\newcommand{\expM}{{\mbox{\tiny M}}}
\newcommand{\uG}{{_\Gamma}}
\newcommand{\uXi}{{_\Xi}}
\newcommand{\uPsi}{{_\Psi}}
\newcommand{\id}{\hat{\mathbb{I}}}

\newcommand{\HH}{{\cal H}}
\newcommand{\proj}[1]{\ket{#1}\!\!\bra{#1}}
\newcommand{\exval}[1]{\langle{#1}\rangle}
\graphicspath{{figures/}}
\begin{document}
\title{Whenever a quantum environment emerges as a classical
system, 
it behaves like a measuring apparatus}
\author{Caterina Foti}
\email{caterina.foti@unifi.it}
\address{Dipartimento di Fisica e Astronomia, Universit\`a di Firenze, I-50019, Sesto Fiorentino (FI), Italy}
\address{INFN, Sezione di Firenze, I-50019, Sesto Fiorentino (FI), Italy}
\author{Teiko Heinosaari}
\address{QTF Centre of Excellence, Turku Centre for Quantum Physics, Department of Physics and Astronomy, University of Turku, 
FIN-20014, Turku, Finland}
\author{Sabrina Maniscalco}
\address{QTF Centre of Excellence, Turku Centre for Quantum Physics, Department of Physics and Astronomy, University of Turku, 
FIN-20014, Turku, Finland}
\author{Paola Verrucchi}
\address{ISC-CNR, at Dipartimento di Fisica e Astronomia, Universit\`a di Firenze, I-50019, Sesto Fiorentino (FI), Italy}
\address{Dipartimento di Fisica e Astronomia, Universit\`a di Firenze, I-50019, Sesto Fiorentino (FI), Italy}
\address{INFN, Sezione di Firenze, I-50019, Sesto Fiorentino (FI), Italy}
\date{\today}
\begin{abstract} 
We study the dynamics of a quantum system $\Gamma$
with an environment $\Xi$ made of $N$ elementary quantum 
components. We aim at answering the following questions:
can the evolution of $\Gamma$ be characterized by some general features
when $N$ becomes very large, 
regardless of the specific form of its interaction with each and every 
component of $\Xi$? In other terms: should 
we expect all quantum systems with a macroscopic environment to undergo 
a somehow similar evolution? And if yes, of what type?
In order to answer these questions we use well established results 
from large-$N$ quantum field theories, particularly referring to the 
conditions ensuring a large-$N$ quantum model to be effectively 
described by a classical theory. 
We demonstrate that the fulfillment of these conditions, when properly 
imported into the framework of the open quantum systems dynamics,  
guarantees that the evolution of $\Gamma$ is always of the same 
type of that expected if $\Xi$ were a measuring apparatus, 
no matter the details of the actual interaction. On the other 
hand, such details are found to determine the specific basis w.r.t. which 
$\Gamma$ undergoes the decoherence dictated by the dynamical 
description of the quantum measurement process.
This result wears two hats: on the one hand it clarifies the physical
origin of the formal statement that, under certain conditions, any
channel from $\rho_\Gamma$ to $\rho_\Xi$ takes the form of a
measure-and-prepare map, as recently
shown in Ref.~\cite{BrandaoPH15};
on the other hand, it formalizes the qualitative argument that
the reason why we do not observe state superpositions is the continual
measurement performed by the environment.
\end{abstract}
\maketitle
\section{Introduction}
\label{s.Introduction}

There exist two closely-related questions about the quantum mechanical 
nature of our universe that keep being intriguing after decades of 
thought processing: how is it that we do not experience state
superpositions, and why we cannot even see them when observing 
quantum systems. 
As for the latter question, it is somehow assumed that this is due to 
the continual measurement process acted upon by the environment. 
However, despite often being considered as an acceptable answer, this 
argument is not a formal result, and attempts to make it such have been only 
recently proposed ~\cite{BrandaoPH15,Zurek09,HorodeckiKH15}. In fact, the current 
analysis of the quantum measurement process~\cite{BuschEtal16}, its 
Hamiltonian description ~\cite{Ozawa1984,LiuzzoScorpoCV15epl}, as well as its 
characterization in the framework of the open quantum systems (OQS) 
dynamics~\cite{HeinosaariZ12} has revealed the qualitative nature of the 
above argument, thus making it ever more urgent to develop a rigorous 
approach to the 
original question. This is the main goal of our work.

Getting back to the first question, the answer offered by the statement 
that microscopic systems obey quantum rules while macroscopic objects 
follow the classical ones, is by now considered unsatisfactory. 
Macroscopic objects, indeed, may exhibit a distinctive quantum 
behaviour (as seen for instance in superconductivity, Bose-Einstein 
condensation, magnetic properties of large molecules with $S=1/2$), 
meaning that the 
large-$N$ condition is not sufficient per-s\'e  for a system made of $N$ 
quantum particles to behave classically. In fact, there exist assumptions which 
single out the minimal structure any quantum theory should possess if it is 
to have a classical limit~\cite{Yaffe82}. Although variously expressed 
depending on the approach adopted by different authors (see the thorough 
discussion on the relation between large-$N$ limits and classical 
theories developed in Sec.VII of Ref.~\cite{Yaffe82}), these assumptions 
imply precise physical constraints on the quantum theory that describes 
a macroscopic quantum system if this has to behave classically.  
In what follows, these assumptions will formally characterize the 
quantum environment, in order to guarantee that the environment, and it alone, behaves 
classically. The relevance of the sentence "and it alone" must be 
stressed: indeed, the work done in the second half of the last century 
on the $N\to\infty$ limit of quantum theories is quite comprehensive but it 
neglects the case when the large-$N$ system is the big partner of a 
principal quantum system, that only indirectly experiences such limit. 
This is, however, an exemplary situation in quantum 
technologies and OQS, hence the questions asked at the beginning of this 
Introduction have recently been formulated in the corresponding framework 
~\cite{BrandaoPH15,Zurek09,HorodeckiKH15,BraunHS01,ChiribellaDA06,GalveZM16,GiorgiGZ15,
RigovaccaEtal15,KnottEtal18,KorbiczEtal17,PleasanceG17}.

In this work, we develop an original approach which uses results for the 
large-$N$ limit of quantum theories in the framework of OQS dynamics. 
This allows us to show that details of the interaction between a quantum 
principal system $\Gamma$ and its environment $\Xi$ are irrelevant in 
determining the main features of the state of $\Xi$ at any time $\tau$ 
in the large-$N$ limit, as long as such limit implies a classical 
behaviour for $\Xi$ itself. If this is the case, indeed, such state can 
always be recognized as that of an apparatus that measures some 
observable of the principal system. The relation between our findings 
and the two questions that open this section is evident.

The paper is structured as follows. In the first section we define the dynamical maps 
characterizing the two evolutions that we aim at comparing. We do so through 
a parametric representation introduced in Sec.~\ref{s.coherent}. 
In Sec.~\ref{s.largeN}, we focus on a peculiar property of generalized coherent states, 
particularly relevant when the large-$N$ limit is considered. As the environment is 
doomed to be macrocopic and behave classically, we then implement such limit in 
Sec.~\ref{s.macroclassic_env}, being finally able to show what we were looking for. 
In Sec.~\ref{s.discussion} we comment on the assumptions made, 
while the results obtained are summed up in the concluding section.

\section{Schmidt decomposition and dynamical maps}
\label{s.Schmidt}

We consider the unitary evolution of an isolated bipartite system 
$\Psi=\Gamma+\Xi$, with Hilbert space $\HH_\uG\otimes\HH_\uXi$; being 
$\Psi$ isolated, it is
\begin{equation} \ket{\Psi(t)}=e^{-i\hat H t}\ket{\Psi}\;, 
\label{e.true_evo} 
\end{equation} 
where $\hbar=1$ and $\hat H$ is any Hamiltonian, describing whatever
interaction between $\Gamma$ and $\Xi$. 
The state $\ket{\Psi}$ is assumed separable
\begin{equation}
\ket{\Psi}=\ket{\Gamma}\otimes\ket{\Xi}~, 
\label{e.initial_state}
\end{equation}
meaning that we begin 
studying the evolution at a time $t=0$ when both $\Gamma$ and $\Xi$ are 
in pure states. This is not a neutral assumption, and we will
get back to it in Sec.~\ref{s.discussion}.

At any fixed time $\tau$, there exists a 
Schmidt decomposition of the state \eqref{e.true_evo},  
\begin{equation}\
\ket{\Psi(\tau)}=\sum_{\gamma}c_\gamma
\ket{\gamma}\ket{\xi_{\gamma}},
\label{e.schmidt_dec}
\end{equation}
with $\gamma=1,...,{\rm dim}{\cal{H}}_\Gamma$, 
$c_\gamma\in\mathbb{R}^{+}$ for $\gamma\le\gamma_{\rm max}\le {\rm 
dim}{\cal{H}}_\Gamma$, $c_\gamma=0$ for $\gamma>\gamma_{\rm max}$, 
$\sum_{\gamma} c_{\gamma}^2=1$, and the symbol $\otimes$ understood (as 
hereafter done whenever convenient). 
The states $\lbrace \ket{\gamma}\rbrace_{\HH_\Gamma}$, and 
$\lbrace \ket{\xi_j}\rbrace_{\HH_\Xi}$ with 
$j=1,...\dim\HH_\Xi$, form what we will hereafter call the 
$\tau$-Schmidt bases, to remind that the Schmidt decomposition is 
state-specific and therefore depends on the time $\tau$ appearing 
in the LHS of Eq.\eqref{e.schmidt_dec}, in whose RHS we have instead
understood the $\tau$-dependence of $c_\gamma$, $\ket{\gamma}$, and
$\ket{R_\gamma}$, for the sake of a lighter notation.
Consistently with the idea that $\Xi$ is a macroscopic system, we take 
$\gamma_{\rm max}<{\rm dim}{\cal H}_\Xi$: therefore, the states 
$\lbrace\ket{\xi_{\gamma}}\rbrace_{\HH_{\Xi}}$ 
entering Eq.\eqref{e.schmidt_dec} are a subset of the pertaining 
$\tau$-Schmidt basis.
Given that $\ket{\Gamma}$ is fully generic, the unitary evolution 
\eqref{e.true_evo} defines, via $\rho_\uXi={\rm Tr}_\uG \rho_\uPsi$, the 
CPTP linear map (from $\Gamma$- to $\Xi$-states)
\begin{equation} 
{\cal{E}}{:}\proj{\Gamma}\rightarrow\rho_\uXi=\sum_\gamma 
c^2_\gamma\proj{\xi_\gamma}~.
\label{e.true_map} 
\end{equation}
Being the output $\rho_\Xi$ a convex sum of orthogonal 
projectors, Eq.\eqref{e.true_map} might describe a projective 
measurement acted upon by $\Xi$ on the principal system $\Gamma$, 
by what is often referred to as measure-and-prepare (m\&p) map.
However, for this being the case, the {\it probability reproducibility 
condition}~\cite{BuschLM96} must also hold, meaning that, given
\begin{equation}
\ket{\Gamma}=\sum_\gamma a_\gamma\ket{\gamma}~,
\label{e.initial_Gamma}
\end{equation}
it should also be $c^2_\gamma=|a_\gamma|^2, \forall \gamma$. This condition, however, 
cannot be generally true, if only for the $\tau$-dependence of the 
Schmidt coefficients $\{c_\gamma\}$ which is not featured by the set 
$\{a_\gamma\}$. In fact, there 
exists a dynamical model (the Ozawa's model~\cite{Ozawa1984} for 
projective von Neumann measurement described in Appendix \ref{a.Ozawa})
for which $c_\gamma^2=|a_\gamma|^2, \forall\gamma$ and $\forall\tau$.
Such model is defined by a 
Hamiltonian where the operators acting on $\Gamma$ must commute with 
each other, a condition that identifies what we will hereafter dub a 
measure-like Hamiltonian, $\hat H^\expM$, with the apex M 
hinting at the corresponding measurement process. The 
evolution defined by $\exp\{-it\hat H^\expM\}$ 
will be consistently dubbed measure-like dynamics
\footnote{Giving a Hamiltonian description of more general quantum 
measurement processes, i.e., identifying the appropriate propagator for 
the dynamics of such processes up to the output production, is a very 
relevant problem that has recently attracted the interest of several 
authors, including some of us.}.

Once established that Eq.\eqref{e.true_map} does not define a 
m\&p map, we can nonetheless use the elements provided by the 
Schmidt decomposition as ingredients to construct a measure-like 
Hamiltonian $\hat H^\expM$ whose corresponding m\&p map, ${\cal 
E}^\expM:\proj{\Gamma}\to\rho_\Xi^\expM$ is the "nearest" possible to 
the actual ${\cal E}$, Eq.\eqref{e.true_map}.

To this aim, we first use the $\tau$-Schmidt bases $\{\ket{\gamma}\}_{\HH_\Gamma}$ 
and $\{\ket{\xi_j}\}_{\HH_\Xi}$ to define the hermitian 
operators
\begin{equation}
\hat O_\uG=\sum_{\gamma}\varepsilon_{\gamma}\proj{\gamma}~~~,~~~
\hat O_\uXi=\sum_jE_j\proj{\xi_j}~,~
\label{e.Schmidt-operators}
\end{equation}
with $\varepsilon_\gamma,E_j$ arbitrary real numbers; we then write the 
interaction Hamiltonian
\begin{equation}
\hat H^{\mbox{\tiny M}}=g\hat O_\uG\otimes\hat O_\uXi\,,
\label{e.meas_H}
\end{equation}
with $g$ some coupling constant, which has the form 
prescribed by the Ozawa's model (see Appendix \ref{a.Ozawa} for more 
details).

Further using the Schmidt coefficients, we construct the separable state
\begin{equation}\label{e.meas_state_in}
\ket{\Psi^{\mbox{\tiny M}}}=\ket{\Gamma}\otimes\ket{\Xi^{\mbox{\tiny 
M}}}\;,
\end{equation}
where $\ket{\Gamma}$ is the same as in Eq.\eqref{e.initial_state}, while 
$\ket{\Xi^{\mbox{\tiny M}}} 
=\sum_\gamma c_\gamma\ket{\xi_\gamma}~,$ with
$c_\gamma$ and $\ket{\xi_\gamma}$ as in Eq.\eqref{e.schmidt_dec}.
Finally we define
\begin{equation}
\ket{\Psi_\tau^{\mbox{\tiny M}}}= e^{-i\hat H^{\mbox{\tiny M}}\tau}
\ket{\Psi^{\mbox{\tiny M}}},
\label{e.meas_evo_def}
\end{equation}
that reads, using $\hat O_\uG\ket{\gamma}=\varepsilon_\gamma\ket{\gamma}$,
$\hat O_\uXi\ket{\xi_{\gamma}}=E_{\gamma}\ket{\xi_{\gamma}}$, and 
$\ket{\Gamma}=\sum_\gamma a_\gamma\ket{\gamma}$,
\begin{align}
\ket{\Psi_\tau^{\mbox{\tiny M}}}=
& e^{-i\hat H^{\mbox{\tiny M}}\tau}
\sum_{\gamma} a_{\gamma}\ket{\gamma}
\sum_{\gamma'}c_{\gamma '}\ket{\xi_{\gamma '}}
\nonumber\\
=& \sum_{\gamma,\gamma'}a_{\gamma}\ket{\gamma}c_{\gamma'}
e^{-i\varphi_{\gamma\gamma'}}
\ket{\xi_{\gamma'}}\;,
\label{e.meas_evo}
\end{align}
with $\varphi_{\gamma\gamma'}\equiv\tau g\varepsilon_\gamma 
E_{\gamma'}\in\mathbb{R}$. Do notice the different notation for the 
time-dependence in Eqs.~\eqref{e.schmidt_dec} and 
\eqref{e.meas_evo_def}: this is to underline that while the former 
indicates how the state $\ket{\Psi}$ of a system with Hamiltonian 
$\hat{H}$ evolves into $\ket{\Psi(t)}$ at any time $t$,
the latter represents a state whose dependence on
$\tau$ not only enters as a proper time in the propagator, but also, as
a parameter, in the definition of $\hat{H}^\expM$ and $\ket{\Xi^\expM}$, 
via the $\tau$-dependence of the Schmidt decomposition 
\eqref{e.schmidt_dec}. Nonetheless, the state $\ket{\Psi^\expM_\tau}$ 
can still be recognized as that in which $\Psi$ would be at time $\tau$, 
were its initial state $\ket{\Psi^\expM}$ and its evolution ruled by the 
measure-like interaction Eq.~\eqref{e.meas_H}.

Given that $\ket{\Gamma}$ is fully generic,
Equation \eqref{e.meas_evo_def} defines, via 
$\rho_\uXi={\rm Tr}_\uG \rho_\uPsi$, the CPTP map from 
$\Gamma$- to $\Xi$-states
\begin{align}
{\cal{E}^\expM}&:\proj{\Gamma}\rightarrow\rho_\uXi^\expM=
\nonumber\\
&=\sum_{\gamma \gamma'\gamma''}|a_\gamma|^2 c_{\gamma'} c_{\gamma''}
e^{i(\varphi_{\gamma\gamma''}-\varphi_{\gamma\gamma'})}
\ket{\xi_{\gamma'}}\!\!\bra{\xi_{\gamma''}}~.
\label{e.meas_map}
\end{align}
Notice that ${\cal E}^{\rm M}$ depends on $\tau$ directly,
via $\varphi_{\gamma\gamma'}\propto\tau$, and indirectly, via the $\tau$-dependence 
of the Schmidt decomposition, that is of the
coefficients $c_\gamma$ and the states $\ket{\xi_\gamma}$.
Comparing Eqs.\eqref{e.true_map} and \eqref{e.meas_map} we 
see that ${\cal E}^\expM$ has the right coefficients 
$\{|a_\gamma|^2\}$ but the wrong form, i.e., it is not a sum of 
orthogonal projectors, while ${\cal E}$ has the correct form but with 
the wrong coefficients, $\{c_\gamma^2\}$. 
In fact, were these two maps equal in some limit, it 
would mean the following: for each time $\tau$, there exists an observable for $\Gamma$, 
(depending on $\tau$ itself) such that the state into which $\Xi$ has 
evolved due to its true interaction with $\Gamma$ is the same, 
in such limit, as if $\Xi$ itself were  
some measuring apparatus proper to that observable, which is quite a statement. 
Since ${\cal{E}}$ and ${\cal{E}^\expM}$ are linear, 
they are the same map iff the output states $\rho_\uXi$ and 
$\rho_\uXi^\expM$ are equal for whatever input $\ket{\Gamma}$. We can 
therefore concentrate upon the structure of such output states, which we will 
do in the next section by introducing a proper parametric representation.

\section{Parametric representation with environmental coherent states}
\label{s.coherent}

The parametric representation with environmental coherent states (PRECS) 
is a theoretical tool that has been recently 
introduced~\cite{CCGV13pnas,TesiDario} to specifically address those 
bipartite quantum systems where one part, on its own made by $N$ 
elementary components, shows an emerging classical behaviour 
in the large-$N$ limit 
~\cite{LiuzzoScorpoCV15epl,CalvaniEtal13,LiuzzoScorpoCV15ijtp,FotiCV16,RossiEtal17}. 
The method makes use of generalized coherent states (GCS) for the system 
intended to become macroscopic. 

The construction of GCS, sometimes referred to as group-theoretic, goes 
as follows~\cite{ZhangFG90}.
Associated to any quantum system there is a Hilbert space ${\cal H}$ and a 
dynamical group ${\cal G}$, which is 
the group containing all the propagators that describe possible
evolutions of the system (quite equivalently, ${\cal G}$ is the group
corresponding to the Lie algebra $\mathfrak{g}$ to which all  
the physical Hamiltonians of the system belong).
Once these ingredients are known, a reference 
state $\ket{0}$ is arbitrarily chosen in $\cal{H}$ and the subgroup 
${\cal F}$ of the propagators that leave such state unchanged (apart 
from an irrelevant overall phase) is determined. This is 
usually referred to as the stability subgroup. 
Elements $\hat\omega$ of ${\cal G}$ that do not belong to such 
subgroup, 
$\hat{\omega}\in{\cal G}/{\cal F}$, generate the GCS 
upon acting on the reference state, $\hat\omega\ket{0}=\ket{\omega}$, 
and are usually dubbed "displacement" operators. 
The GCS construction further entails the definition of an 
invariant\footnote{The measure $d\mu(\hat\omega)$ is called \emph{invariant} 
because it is left unchanged by the action of ${\cal 
G}$.}
measure $d\mu(\hat\omega)$ on ${\cal G}/{\cal F}$ such that a resolution 
of the identity on ${\cal{H}}$ is provided in the form
\begin{equation}
\int_{{\cal G}/{\cal F}}d\mu(\hat\omega)\proj{\omega}=\id_{\cal{H}}~.
\label{e.ECS_resolid}
\end{equation}

One of the most relevant byproduct of the GCS construction is the
definition of a differentiable manifold ${\cal{M}}$ via the chain of
one-to-one correspondences
\begin{equation}
\hat{\omega}\subset{\cal G}/{\cal F}\Leftrightarrow\ket{\omega}\in{\cal 
H}\Leftrightarrow\omega\subset{\cal M}~,
\label{e.GCScorrespondence}
\end{equation}
so that to any GCS is univoquely associated a point on ${\cal M}$, and 
viceversa. A measure $d\mu(\omega)$ on ${\cal 
M}$ is consistently associated to the above 
introduced $d\mu(\hat\omega)$, so that requiring GCS to be normalized, 
$\exval{\omega|\omega}=1$, implies
\begin{align}
\exval{\omega|\omega}&=\exval{\omega|\left[ \int_{{\cal G}/{\cal F}}
d\mu(\hat\omega)\proj{\omega}\right]|\omega}\nonumber\\
&=\int_{\cal 
M}d\mu(\omega)|\exval{\omega|\omega}|^2=1~;
\label{e.GCS_non-orthogonal}
\end{align}
notice that GCS are not necessarily orthogonal.

One important aspect of the GCS construction is that it
ensures the function $\exval{\omega|\rho|\omega}$ for whatever state 
$\rho$ (often called Husimi 
function in the literature\footnote{In fact, a "Husimi function" is in 
principle defined on a classical phase-space, while ${\cal M}$ is a 
differential manifold with a simplectic structure that should not be 
considered a phase-space, yet, i.e., before the large-$N$ limit is taken;
however, it is quite conventional to extend the term to the expectation 
value of $\rho$ on GCS.})
is a well-behaved probability 
distribution on $\cal M$ that uniquely 
identifies $\rho$ itself. As a consequence, 
studying 
$\exval{\omega|\rho|\omega}$ on $\cal M$ is fully equivalent to perform 
a state-tomography of $\rho$ on the Hilbert space, and 
once GCS are available one can analyze any state $\rho$ 
of the system by studying its Husimi function on $\cal M$, which is what 
we will do in the following.
We refer the reader to Refs.~\cite{ZhangFG90,Perelomov72} for 
more details.

When GCS are relative to a system $\Xi$ which is 
the environment of a principal system $\Gamma$, we call them 
Environmental Coherent States (ECS). 

\begin{figure}
\centering
\includegraphics[width=.45\textwidth]{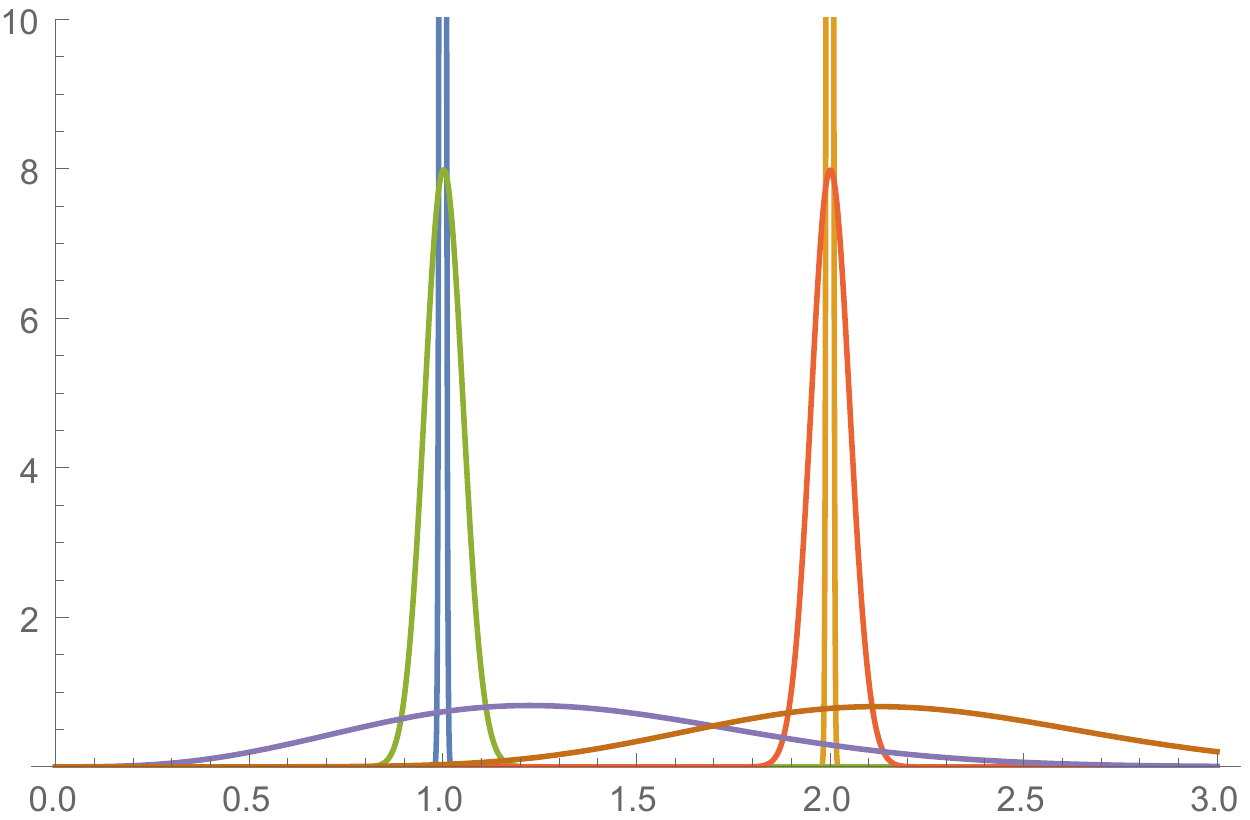}
\caption{ $|\exval{\alpha|n}|^2$ as a function of 
$\sqrt{\alpha\alpha^*}$, with $n=1$ (left) and 
$n=4$ (right), for $N=1,10,1000$ (bottom to top).}
\label{f.bosonoverlap1}
\end{figure}
\begin{figure*}
\centering
\subfigure{\includegraphics[width=.30\textwidth]{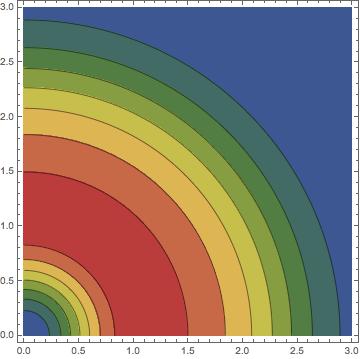}},
\subfigure{\includegraphics[width=.30\textwidth]{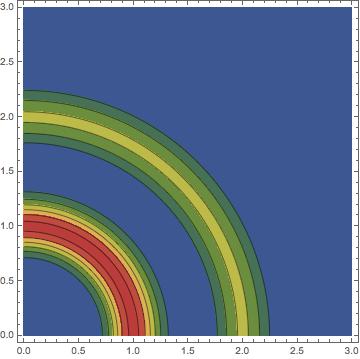}}'
\subfigure{\includegraphics[width=.30\textwidth]{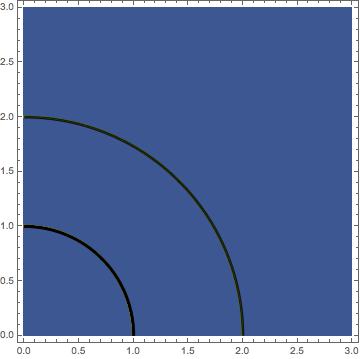}}
\caption{Sum $|\exval{\alpha|n'}|^2+|\exval{\alpha|n{''}}|^2$ with 
$n'=1$ and $n{''}=4$ for 
$N=1,10,1000$ 
(left to right): Contourplot on part of ${\cal M}$, which is now the 
complex plane (values increase from blue to red).}
\label{f.bosonoverlap2}
\end{figure*}

Getting back to the setting of section~\ref{s.Schmidt}, we first 
recognize that, if they were to represent different evolutions of the 
same physical system, the propagators $\exp\{-i\hat H\tau\}$ and 
$\exp\{-i\hat{H}^\expM\tau\}$ must belong to the same dynamical group, 
as far as their action on ${\cal H}_\uXi$ is concerned. 
More explicitely, this group is identified as follows:
$i)$ consider all the operators acting on ${\cal H}_\Xi$ in the total 
Hamiltonians $\hat H$ and $\hat H^M$;
$ii)$ find the algebra to which they all belong
(notice that, as both Hamiltonians refer to the same physical system,
the above operators must belong to the same algebra $\mathfrak{g}$;
$iii)$ recognize the dynamical group as that associated to the above 
algebra $\mathfrak{g}$ via the usual exponential Lie map (for several examples see 
for instance Refs.~\cite{LiuzzoScorpoCV15epl,CCGV13pnas,FotiCV16,RossiEtal17}).
This is the 
group to be used for constructing the ECS, according to the procedure 
briefly sketched above. Once ECS are constructed, the PRECS of any 
pure state $\ket{\psi}$ of 
$\Psi$ is obtained by inserting an identity resolution
in the form \eqref{e.ECS_resolid} into any decomposition of 
$\ket{\psi}$ as linear combination of separable (w.r.t. the 
partition $\Psi=\Gamma+\Xi$) states.
Explicitly, one has
\begin{equation}
\ket{\psi}=\int_{\cal M} 
d\mu(\omega)\chi(\omega)\ket{\omega}\ket{\Gamma(\omega)}~,
\label{e.PRECS_general}
\end{equation}
where $\ket{\Gamma(\omega)}$ is a normalized state for $\Gamma$ that 
parametrically depends on $\omega$, while $\chi(\omega)$ is a real 
function on $\cal M$ whose square 
\begin{equation}
\chi(\omega)^2=\exval{\omega|\rho_\uXi|\omega}~,
\label{e.Husimi}
\end{equation}
is the environmental Husimi function relative to 
$\rho_\uXi={\rm Tr}_\uG\proj{\psi}$, 
i.e., the normalized distribution on ${\cal M}$ 
that here represents the probability for the environment $\Xi$ to be in 
the GCS $\ket{\omega}$ when $\Psi$ is in the pure state 
$\ket{\psi}$. 
The explicit form of $\chi(\omega)$ and $\ket{\Gamma(\omega)}$ is 
obtained from any decomposition of $\ket{\psi}$ into a linear 
combination of separable (w.r.t. the partition $\Gamma+\Xi$) states.

In particular,
for the states 
\eqref{e.schmidt_dec} and \eqref{e.meas_evo}, it is
\begin{equation}
\chi(\omega)^2=
{\sum_{\gamma}c_\gamma^2 
\left| \bra{\omega}\xi_\gamma\rangle \right|^2}~,
\label{e.chi2_true}
\end{equation}
and
\begin{align}
&\chi^\expM(\omega)^2=\nonumber\\
&=\sum_{\gamma\gamma'\gamma''}|a_{\gamma}|^2 
c_{\gamma'} 
c_{\gamma''}e^{i(\varphi_{\gamma\gamma''}-\varphi_{\gamma\gamma'})}
\bra{\omega}\xi_{\gamma'}\rangle
\bra{\xi_{\gamma''}}\omega\rangle~,\nonumber\\
&~
\label{e.chi2_meas}
\end{align}
respectively.

Comparing $\chi(\omega)^2$ and $\chi^\expM (\omega)^2$ is equivalent to 
compare
$\rho_\Xi$ and $\rho_\Xi^\expM$, and hence the maps 
\eqref{e.true_map} and \eqref{e.meas_map}.
However, despite the very specific construction leading 
to $\ket{\Psi^\expM_\tau}$, we cannot yet make any meaningful specific  
comparison between 
$\chi(\omega)^2$ and $\chi^\expM(\omega)^2$ at this stage. 
Indeed, we still have to exploit the fact that the 
environment is doomed to be big and behave classically, which is why ECS 
turn out to be so relevant to the final result, as shown in the next section.

\section{Large-$N$ and classical limit}
\label{s.largeN}

As mentioned in the Introduction, a physical system which is made by a 
large number $N$ of quantum constituents does not necessarily obey the 
rules of classical physics. However, several authors 
~\cite{Yaffe82,ZhangFG90,Lieb73,GnutzmannK98} have shown 
that if GCS exist and feature some 
specific properties, then the structure of a classical theory ${\cal C}$ emerges
from that of a quantum theory ${\cal Q}$.
In particular, the existence of GCS establishes a relation between the 
Hilbert space of ${\cal Q}$ and the manifold ${\cal M}$ that their 
construction implies, which turns out to be the phase-space of the 
classical theory that emerges as the large-$N$ limit of ${\cal Q}$. 
In fact, one should rather speak about the $k\to 0$ limit 
of ${\cal Q}$, with $k$ the real positive number, referred to
as "quanticity parameter", such that all the commutators of 
the theory (or anticommutators, in the fermionic case) 
vanish with $k$. However, all known quantum theories for systems 
made by $N$ components have $k\sim \frac{1}{N^p}$
with $p$ a positive number: therefore, for the sake of clarity, we 
will not hereafter use the vanishing of the quanticity parameters but 
rather refer to the large-$N$ limit (see Appendix \ref{a.large_N} 
for more details).

Amongst the above properties of GCS, that are thoroughly explained and 
discussed in Ref.~\cite{Yaffe82} as the {\it assumptions} guaranteeing the large-$N$ 
limit to define a classical theory, one that plays a key role in 
this work regards the overlaps $\exval{\omega|\xi}$, whose square 
modulus represents the probability that a system in some generic pure 
state $\ket{\xi}$ be observed in the coherent state $\ket{\omega}$.
These overlaps never vanish for finite $N$, due to the overcompleteness 
of GCS: as a consequence, if one considers two orthonormal states, say
$\ket{\xi'}$ and $\ket{\xi{''}}$, there might be a finite 
probability for a system in a GCS $\ket{\omega}$ to be observed either 
in $\ket{\xi'}$ or in $\ket{\xi{''}}$.
This formally implies that, defined $S_{\xi}$ the set of points on 
$\cal{M}$ where 
$|\exval{\omega|\xi}|>0$, it generally 
is $S_{\xi '}\cap S_{\xi{''}}\neq\emptyset$. 
\begin{figure*} 
\centering
\subfigure{\includegraphics[width=.30\textwidth]{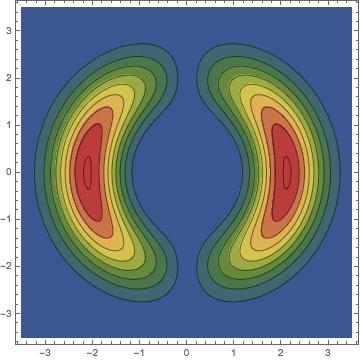}},
\subfigure{\includegraphics[width=.30\textwidth]{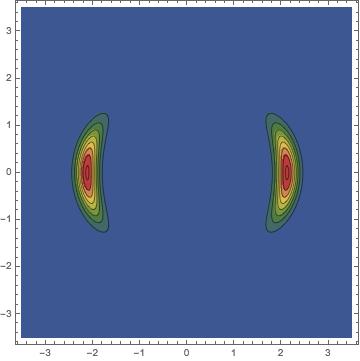}}'
\subfigure{\includegraphics[width=.30\textwidth]{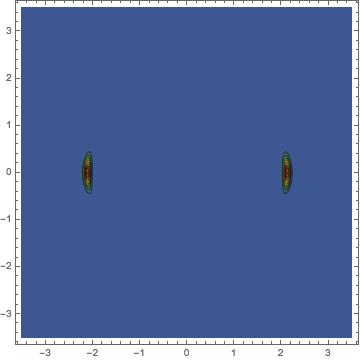}}
\caption{Sum $|\exval{\alpha|+}|^2+|\exval{\alpha|-}|^2$ with 
$\ket{\pm}=(\ket{1}\pm\ket{2})/\sqrt{2}$, for 
$N=1,10,1000$ (left to right): Contourplot on ${\cal M}$, which is now the 
complex plane (values increase from blue to red).}
\label{f.bosonoverlap3}
\end{figure*}

On the other hand, the quantity
\begin{equation}
\lim_{N\to\infty}|\exval{\omega|\xi}|^2
\label{e.clas_lim_generic} 
\end{equation}
features some very relevant properties.
First of all, if $\ket{\xi}$ is another GCS, say 
$\ket{\omega'}$, the square modulus $|\exval{\omega|\omega'}|^2$ exponentially 
vanishes with $|\omega-\omega'|^2$ in such a way that the limit 
\eqref{e.clas_lim_generic} converges to the Dirac distribution 
$\delta(\omega-\omega')$, thus restoring a notion of distinguishability 
between different GCS in the large-$N$ limit.
Moreover, in Appendix \ref{a.large-N_Overlap} we demonstrate that
\begin{equation}
\exval{\xi'|\xi{''}}=\delta_{\xi'\xi{''}}\Leftrightarrow
\lim_{N\to\infty} S_{\xi'}\cap S_{\xi''}=\emptyset~,
\label{e.supports-dont-share}
\end{equation}
meaning that orthonormal states are put together by 
distinguishable sets of GCS. In other terms, the large-$N$ limit 
enforces the emergence of a one-to-one correspondence between elements of any 
orthonormal basis $\{\ket{\xi}\}$ and disjoint sets of GCS, 
in such a way that the distinguishability of the former is reflected 
into the disjunction of the latter.
Given the relevance of Eq.\eqref{e.supports-dont-share} to this work, 
let us discuss its meaning with two explicit examples.

\subsection{Field Coherent States}

Consider a system $\Xi$ whose Lie algebra is $\mathfrak{h}_4$, i.e., the vector 
space spanned by 
$\{\hat a,\hat a^\dagger, \hat n\equiv\hat a^\dagger \hat a, \id\}$, with 
Lie brackets $[\hat a,\hat a^\dagger]=1$, and 
$[\hat a^{(\dagger)},\hat n]=(-)\hat a^{(\dagger)}$.
In order to identify the quanticity parameter $k$, i.e., the parameter 
whose vanishing makes the Lie brackets of the theory go to zero, one can
restore dimensionful ladder operators,
$\hat a^{(\dagger)}\to\sqrt{\frac{2\hbar}{M\omega}}\hat 
a^{(\dagger)}$, and observe that all the commutators vanish in the 
large-$M$ limit. Further taking $M\propto N$, meaning that 
the total mass of $\Xi$ is the sum of the masses of the elementary 
components, which are assumed to have the same mass for the sake of 
simplicity, it is easily found 
that $k\sim 1/N$.
As for the GCS , they are the well known field coherent 
states 
$\{\ket{\alpha}\}$, with $\ket{0}:\hat a\ket{0}=0$ the 
reference state, and ${\cal M}$ the complex plane. The eigenstates of 
$\hat n$ are the Fock states $\{\ket{n}\}$, and 
$\exp\{\alpha \hat a - \alpha^* \hat a^\dagger\}\equiv \hat \alpha$ 
is the displacement operator such that 
$\ket{\alpha}=\hat\alpha\ket{0}$.

As for the overlaps entering Eq.\eqref{e.chi2_meas}, let us 
first consider the case when the states $\{\ket{\xi_\gamma}\}$ are Fock 
states. In Fig.~\ref{f.bosonoverlap1} we show $|\exval{\alpha|n}|^2$ as 
a function of $|\alpha|^2$, for 
$n=1,2$ and different values of $N$. 
It is clearly seen that $S_{n'}\cap S_{n''}\to\emptyset$ as 
$N\to\infty$, meaning that the product of overlaps in Eq.\eqref{e.chi2_meas} 
vanishes unless $\gamma'=\gamma{''}$, i.e. $n'=n{''}$ in this specific example. 
In order to better visualize $S_{n'}$ and $S_{n{''}}$ on ${\cal M}$, in 
Fig.~\ref{f.bosonoverlap2} we contour-plot the sum 
$|\exval{\alpha|1}|^2+|\exval{\alpha|2}|^2$:
indeed we see that, as $N$ increases, $S_1$ and 
$S_2$ do not intersect. Notice that increasing $N$ does not squeeze 
$S_n$ to the neighbourghood of some point on $\cal M$, as is the case for
$\lim_{N\to\infty}|\exval{\alpha|\alpha'}|^2=\delta(\alpha-\alpha')$, 
but rather to that of the circle $|\alpha|^2=n$.
In other terms, more field coherent states overlap with the same Fock 
state, but different Fock states overlap with distinct sets of field 
coherent states, in the large-$N$ limit.
This picture holds not only for Fock states but, as expressed by 
Eq.\eqref{e.supports-dont-share}, for any pair of orthonormal states.
In Fig.~\ref{f.bosonoverlap3}, for instance, we contour-plot the sum
$|\exval{\alpha|+}|^2+\exval{\alpha|-}|^2$ with
$\ket{\pm}\equiv(\ket{1}\pm\ket{2})/\sqrt{2}$: in this case $S_+$ and 
$S_-$ are disjoint already for $N=1$, and keep shrinking as $N$ 
increases.

\subsection{Spin Coherent States}

\begin{figure*} 
\centering
\subfigure{\includegraphics[width=.30\textwidth]{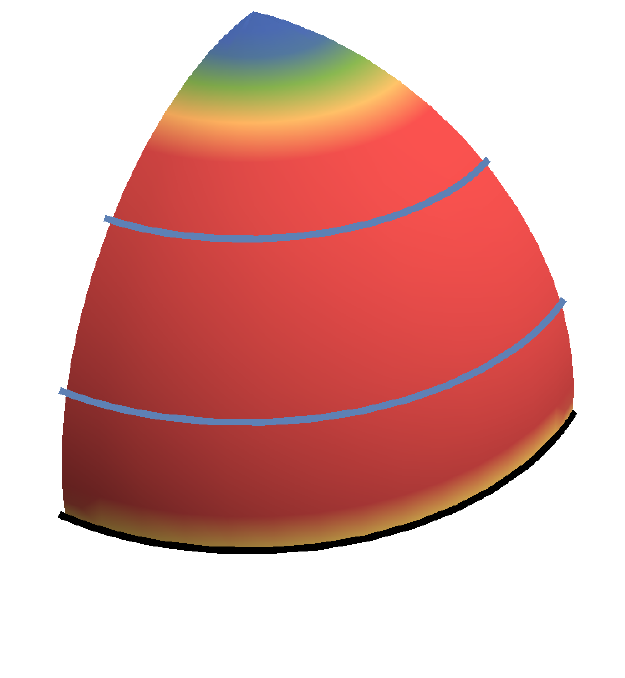}},
\subfigure{\includegraphics[width=.30\textwidth]{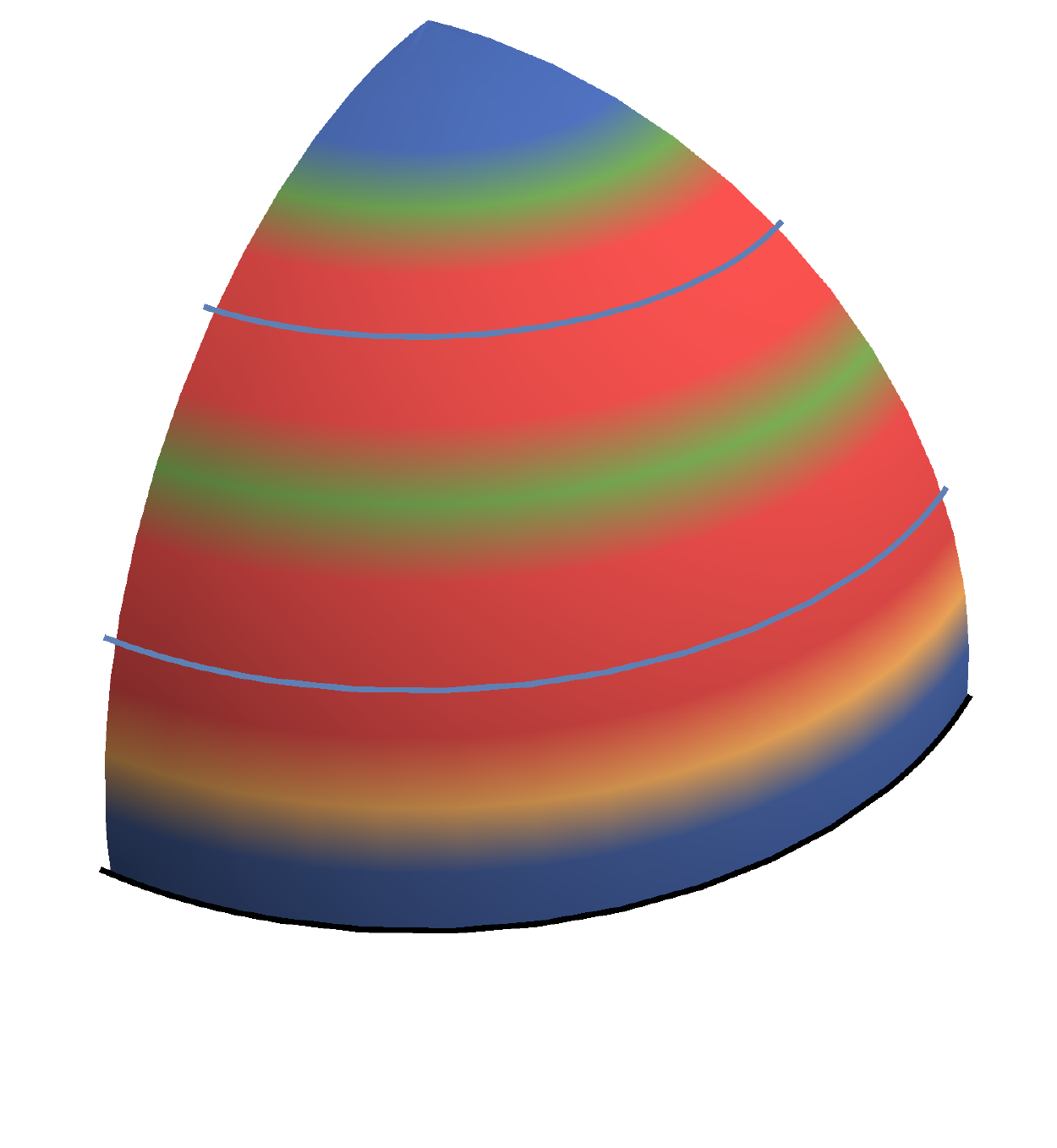}}
\subfigure{\includegraphics[width=.30\textwidth]{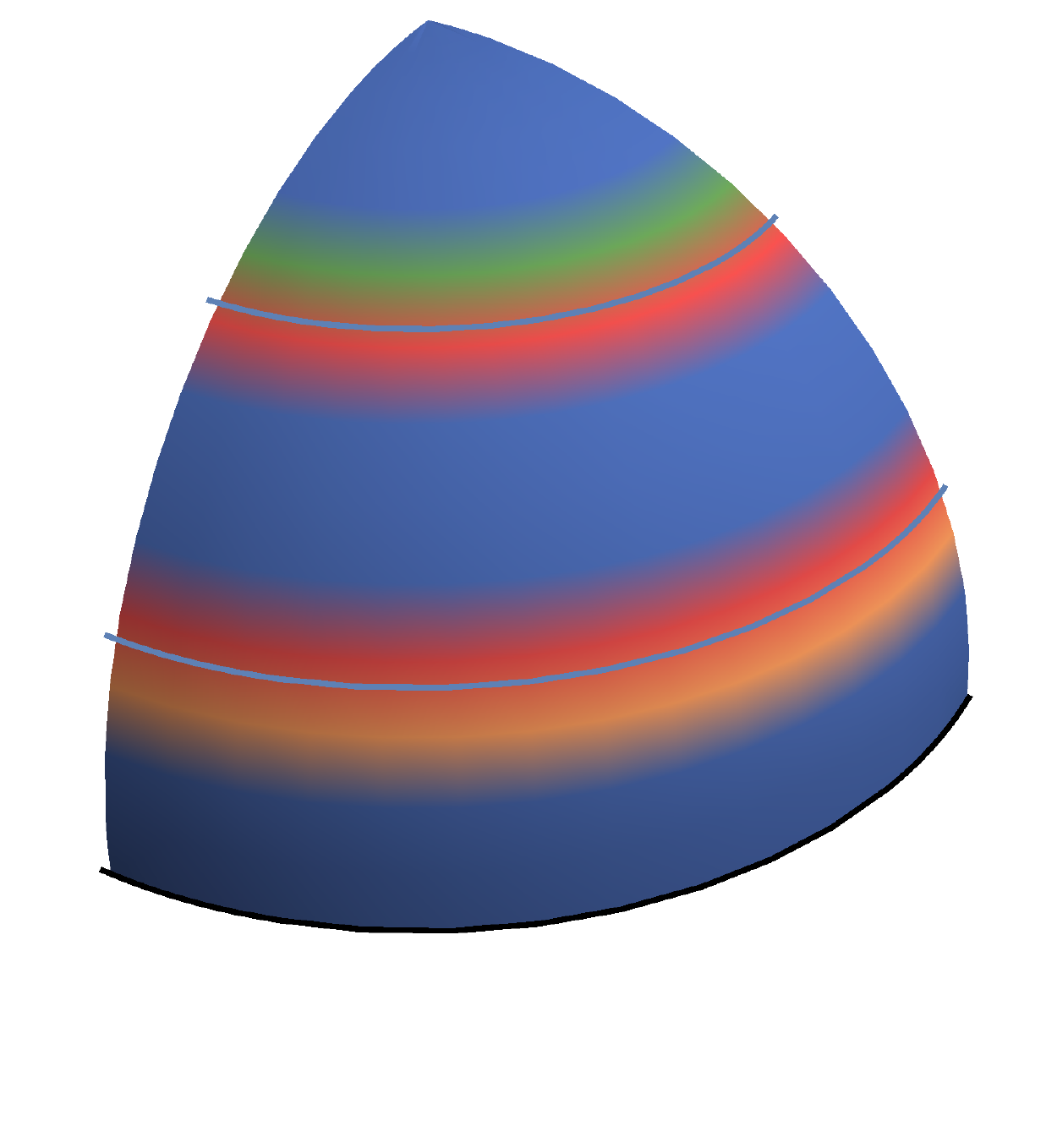}}
\caption{Sum $|\exval{\Omega|m'}|^2+|\exval{\Omega|m"}|^2$ with 
$m'/S=0.8$ and 
$m{''}/S=0.4$, for $N=10,100,1000$ (left to right): Densityplot 
on part of ${\cal M}$, which is now the unit sphere (values increase 
from blue to red).}
\label{f.spinoverlap2}
\end{figure*}

A very similar scenario appears when studying
a system $\Xi$ whose Lie algebra is $\mathfrak{su}(2)$, i.e., the vector space 
spanned by $\{\hat S^+,\hat S^-, \hat S^z\}$, with Lie brackets 
$[\hat S^+,\hat S^-]=2\hat S^z, [\hat S^z,\hat S^\pm]=\pm\hat S^\pm$, 
and $|\hat{\mathbf S}|^2=S(S+1)$, with $S$ fixed and constant; 
in this case the quanticity parameter is 
identified by noticing that the normalized operators 
$\hat s^*\equiv \frac{1}{S}\hat S^*$, $*=z,\pm$, have vanishing commutators in the 
large-$S$ limit. Further taking $S\propto N$, meaning that the 
total spin of $\Xi$ is a conserved quantity, whose value is the sum of the 
spins of each individual component, it is easily found that 
$k\sim 1/N$.
As for the GCS , they are the so-called spin (or atomic) coherent 
states $\{\ket{\Omega}\}$, with the reference state 
$\ket{0}:\hat S^z\ket{0}=-S\ket{0}$, 
and ${\cal M}$ the unit sphere. 
The eigenstates of $\hat S^z$  are
$\{\ket{m}\}:\hat S^z\ket{m}=(-S+m)\ket{m}$, and the 
displacement operators are
$\hat\Omega=\exp\{\eta \hat S^- - \eta^* \hat S^+\}$,
with $\eta=\frac{\theta}{2}e^{i\phi}$, and 
$\theta\in[0,\pi],\phi\in[0,2\pi)$ the spherical coordinates.
\begin{figure}
\centering
\includegraphics[width=.45\textwidth]{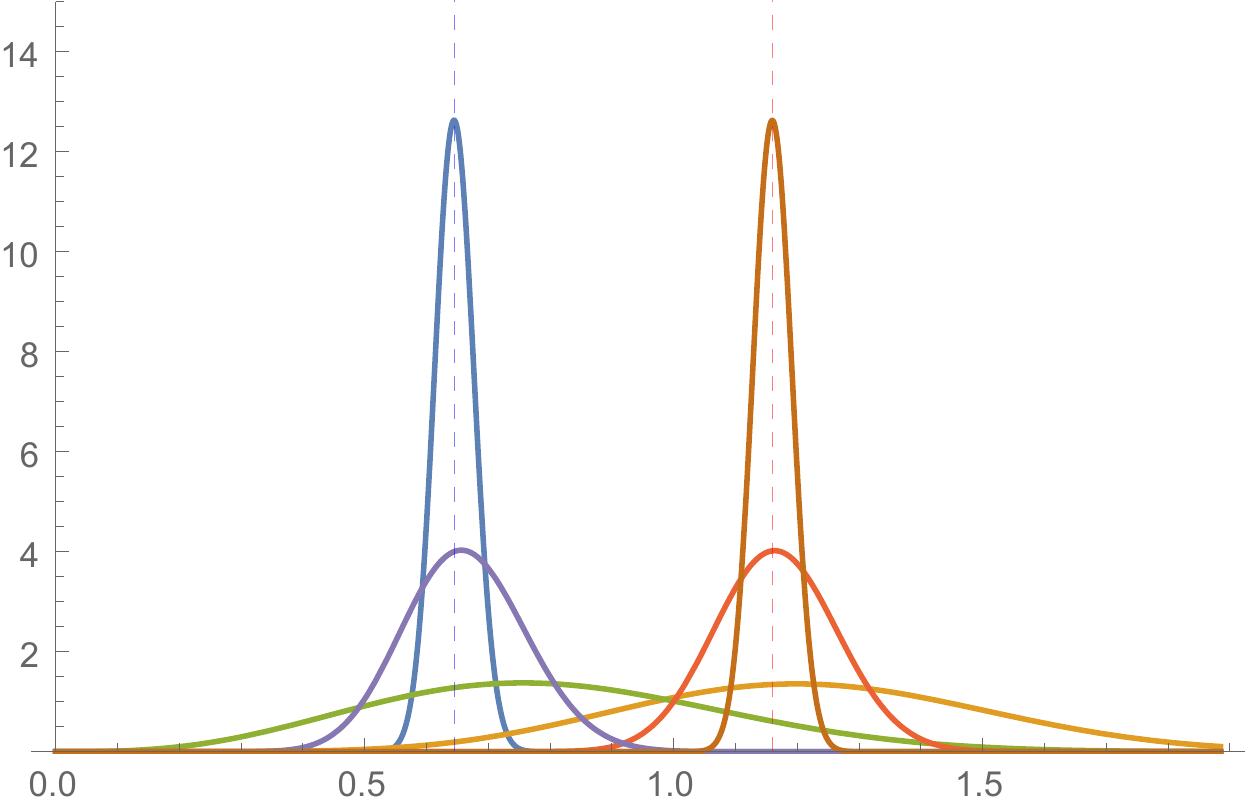}
\caption{ $|\exval{\Omega|m}|^2$ as a function of
$\theta$, for $m/S=0.8$ (left) and
$0.4$ (right), for $N=10,100,1000$ (bottom to top).}
\label{f.spinoverlap1}
\end{figure}
As for the overlaps entering Eq.\eqref{e.chi2_meas},
the analytical expression for $\exval{\Omega|m}$ is available (see for 
instance Ref.~\cite{ZhangFG90}), which allows us to show,
in Fig.~\ref{f.spinoverlap1}, the square modulus
$|\exval{\Omega|m}|^2$ for $m'/S=0.8$ and 
$m{''}/S=0.4$,
for different values of $N$. Again we see that 
$S_{m'}\cap S_{m{''}}\to\emptyset$ as $N\to\infty$, implying that the 
product in Eq.\eqref{e.chi2_meas} 
vanishes unless 
$\gamma'=\gamma{''}$, i.e., $m'=m{''}$ in this specific example. In 
Fig.~\ref{f.spinoverlap2} we show the sum 
$|\exval{\Omega|m'}|^2+|\exval{\Omega|m{''}}|^2$ as density-plot on part 
of the unit sphere: besides the expected shrinking of the regions 
where the overlaps are finite, we notice that, as seen in the bosonic 
case, the support of $\lim_{N\to\infty}|\exval{\Omega|m}|^2$ does not 
shrink into the neighbourghood of a point on the sphere, as is the case for
$\lim_{N\to\infty}|\exval{\Omega|\Omega'}|^2=\delta(\Omega-\Omega')$, 
but rather into that of the parallel $\cos\theta=m/S$.

\section{A macroscopic environment that behaves classically}
\label{s.macroclassic_env}

Let us now get back to the general case and to Eq.\eqref{e.chi2_meas}: the 
states $\ket{\xi_{\gamma'}}$ and 
$\ket{\xi_{\gamma{''}}}$ are othonormal by definition, being elements of the 
$\tau$-Schmidt basis $\{\ket{\xi_j}\}_{{\cal H}_\Xi}$ introduced in 
Sec.\ref{s.Schmidt}.
Therefore Eq.\eqref{e.supports-dont-share} holds, meaning
\begin{equation}
\lim_{N\to\infty}\exval{\omega|\xi_{\gamma'}}\exval{\xi_{\gamma{''}}|\omega}
=
\lim_{N\to\infty}|\exval{\omega|\xi_{\gamma'}}|^2\delta_{\gamma'\gamma{''}}~,
\label{e.product-of-overlaps_limit}
\end{equation}
and hence
\begin{equation}
\lim_{N\to\infty}\chi^\expM(\omega)^2=
\sum_{\gamma\gamma'}|a_\gamma|^2c^2_{\gamma'}\lim_{N\to\infty}
|\exval{\omega|\xi_{\gamma'}}|^2~.
\label{e.lim_chi2_meas}
\end{equation}
Using $\sum_\gamma|a_\gamma|^2=1$, and the swap 
$\gamma'\leftrightarrow\gamma$, we finally obtain
\begin{equation}
\lim_{N\to\infty}\chi^\expM(\omega)^2=\lim_{N\to\infty}\chi(\omega)^2~,
\label{e.lim_chi2}
\end{equation}
which is what we wanted to prove, namely that the
the dynamical maps \eqref{e.true_map} and \eqref{e.meas_map}
are equal when $\Xi$ is a quantum macroscopic system whose 
behaviour can be effectively described classically.

\section{Discussion}
\label{s.discussion}

Aim of this section is to comment upon some specific aspects of our 
results, with possible reference to the way other authors have 
recently tackled the same subject. 
Let us first consider the assumption that the initial state 
\eqref{e.initial_state} of the total system $\Psi=\Gamma+\Xi$ be 
separable. If this is not the case, as it may happen, one must look for the 
different partition $\Psi=A+B$, such that 
$\ket{\Psi}=\ket{A}\otimes\ket{B}$. If this partition is still such that 
the subsystem $B$ is macroscopic and behaves classically, the change is 
harmless and the whole construction can be repeated with $A$ the quantum 
system being observed and $B$ its observing environment.
On the other hand, if the new partition is such that neither $A$ nor $B$ 
meet the conditions for being a classical environment, then the problem 
reduces to the usual one of studying the dynamics of two interacting 
quantum systems, for which any approach based on effective descriptions 
is incongrous, as details of the true Hamiltonian will always be 
relevant. Notice that this analysis is fully consistent with the results 
presented in Ref.~\cite{BrandaoPH15}, which are embodied into 
inequalities whose meaning wears off as dim${\cal H}_B$  diminishes.
The case when $\Psi$ is not initially in a pure state is similarly 
tackled by enlarging $\Psi\to\widetilde{\Psi}$ as much as necessary for 
$\widetilde{\Psi}$ to be in a pure state: a proper choice of a new
partition of $\widetilde{\Psi}$ will follow.

We then want to clarify in what sense the Hamiltonian 
\eqref{e.meas_H} is said to induce a "measure-like dynamics" or, which is 
quite equivalent, the channel \eqref{e.meas_map} to define a 
m\&p map: the quotes indicate that the actual output 
production,
which happens at a certain time according to some process whose nature 
we do not discuss, is not considered and it only enters the description 
via the requirement that the probability for each output is the one 
predicted by Born's rule. To this respect, one might also ask what is
the property of $\Gamma$ which is observed by $\Xi$: this is 
the one represented, in the Ozawa's model, by the operator 
$\hat{O}_\Gamma$,
and it therefore depends on the true evolution via the Schmidt 
decomposition of the evolved state. To put it another way, details of the 
interaction do not modify the measure-like nature of the dynamics in the 
large-$N$ limit, but they do affect what actual measurement is performed 
by the environment.

Let us now discuss possible connections between our results 
and Quantum Darwinism~\cite{Zurek09,BrandaoPH15}. 
As mentioned at the end of Appendix \ref{a.large_N},
a sufficient condition for a quantum theory to 
have a large-$N$ limit which is a classical theory is the existence of 
a global symmetry, i.e., such that its  
group-elements act non-trivially upon the 
Hilbert space of each and every component of the total system $\Xi$ that 
the theory describes.
In fact, few simple examples show
that quantum theories with different global symmetries
can flow into the same classical theory in the large-$N$ 
limit: in other words, echoing L.~G.~Yaffe in Ref.~\cite{Yaffe82}, 
different quantum theories can be "classically equivalent".
If one further argues that amongst classically equivalent quantum theories 
there always exists a free theory, describing $N$ non-interacting 
subsystems, it is possible to show that each macroscopic 
fragment of $\Xi$ can be effectively described as if it were the same
measurement apparatus. Work on this point is in progress, 
based on the quantum de Finetti theorem, results from 
Refs.~\cite{ChiribellaDA06,BrandaoPH15}, and the 
preliminary analysis reported in Refs.~\cite{Querini_Master16,Foti_phd2019}. 
We close this section by mentioning the possible connection between
our description and the way the notion of "objective information" is 
seen to emerge in Ref.~\cite{HorodeckiKH15}: in fact, the idea that there may 
be no quantum-to-classical transition involved in the perception of the 
world around us, that might rather emerge just as a reflection of some 
specific properties of the underlying quantum states, seems to be 
consistent with the discussion reported above, and we believe that 
further investigation on this point might be enlightening.

\section{Conclusions}
\label{s.conclusions}

The idea that the interaction with macroscopic environments causes the 
continual state-reduction of any quantum system is crucial for making 
sense of our everyday experience w.r.t. the quantum description of 
nature. However, the formal analysis of this idea has been 
unsatisfactory for decades, due to several reasons, amongst which we 
underline the following.

Firstly the generality of the above idea implies that assumptions on the 
initial state of the quantum system, and the specific form of the 
interaction with its environment, should not be made. Secondly, formal 
tools must be devised to allow the study of the system-plus-environment 
dynamics in a way that guarantees a genuinely quantum description of the 
system throughout the crossover of the environment towards a classical 
behaviour. Finally, a clean procedure is required to ensure that the 
above crossover takes place when the environment becomes macroscopic, 
i.e., in the large-$N$ limit of the quantum theory that describes it.

In this work, reminding that principal system and environment are dubbed 
$\Gamma$ and $\Xi$, respectively, we have addressed the above three 
issues as follows. As for the first point, the analysis is developed by 
comparing CPTP linear maps from $\Gamma$- to $\Xi$-states, that do not 
depend on the initial state of $\Gamma$ by definition. The considered 
maps, Eqs.~\eqref{e.true_map} and \eqref{e.meas_map}, are defined using 
ingredients provided by the Schmidt decomposition of the 
system-plus-environment evolved state, Eq.~\eqref{e.true_evo}, that 
exists at any time, and whatever the form of the interaction between 
$\Gamma$ and $\Xi$ is. Regarding the second issue, we have used a 
parametric representation of the overall system state, 
Eq.~\eqref{e.PRECS_general}, that resorts to generalized coherent 
states (i.e., coherent states as defined via the group-theoretical 
approach) for describing $\Xi$. This representation, both for its 
parametric nature and the peculiar properties of coherent states when 
the quantum-to-classical crossover is considered, allows us to implement 
the large-$N$ limit for $\Xi$ without making assumptions on $\Gamma$ 
or affecting its quantum character. 
The third point has been tackled by using results from large-$N$ quantum 
field theories: these results provided us with formal conditions that 
generalized coherent states must fulfill, particularly 
Eq.~\eqref{e.Yaffe-assumption_3}, in order to describe a macroscopic 
system that behaves classically.

After this elaboration, we have managed to compare the map defined by 
the true evolution of $\Gamma+\Xi$, Eq.~\eqref{e.true_map}, with that 
corresponding to a measure-and-prepare dynamical process, 
Eq.~\eqref{e.meas_map}, in terms of the difference between probability 
functions entering the parametric representation, 
Eqs.~\eqref{e.chi2_true} and \eqref{e.chi2_meas}. These functions have been 
demonstrated to become equal when the large-$N$ limit defines a 
classical dynamics for $\Xi$.

Overall, our approach allows one to tackle the so-called quantum to 
classical crossover \cite{Schlosshauer07} by a rigorous mathematical 
formulation that provides a physically intuitive picture of the 
underlying dynamical process. In fact, exploiting the most relevant fact that 
not every theory has a classical limit, we 
have shown that any dynamics of whatever OQS 
defines a Hamiltonian model that characterizes its environment
as a measuring apparatus
\textit{if} the conditions ensuring that the above classical 
limit exists and corresponds to a large-$N$ condition upon the 
environment itself are fulfilled.
In other words, \textit{if} some dynamics
emerges in the classical world, it \textit{necessarily} is a 
measure-like one.

Let us conclude by briefly commenting upon the already mentioned 
phenomenon known as Quantum Darwinism, introduced in \cite{Zurek09} and 
recently considered in \cite{BrandaoPH15} from an information theoretic 
viewpoint. Our work suggests that Quantum Darwinism might emerge as a 
dynamical process, with its generality due to the versatilility of the 
Hamiltonian model for the quantum measurement process, and the loss of 
resolution inherent in the classical description.

\begin{acknowledgments} CF acknowledges M. Piani and M. Ziman for useful 
and stimulating discussions. SM and TH acknowledge financial support 
from the Academy of Finland via the Centre of Excellence program 
(Project no. 312058) as well as Project no. 287750. CF and PV acknowledge 
financial support from the University of Florence in the framework of the University 
Strategic Project Program 2015 (project BRS00215). 
PV acknowledges 
financial support from the Italian National Research Council (CNR) via
the "Short term mobility" program STM-2015, and declares to have worked 
in the framework of the Convenzione Operativa between the Institute for 
Complex Systems of CNR and the Department of Physics and Astronomy of 
the University of Florence. Finally, CF and PV warmly thank the Turku Centre 
for Quantum Physics for the kind hospitality. \\
\end{acknowledgments}


\onecolumn\newpage
\appendix
\section{From Ozawa's model to the measure-and-prepare map}
\label{a.Ozawa}
Given a projective measurement with measurement operators 
$\{\proj{\pi}\}$ acting on ${\cal{H}}_\Gamma$, its dynamical 
description according to the Ozawa's model is defined by the propagator 
$\exp\{-it\hat H^\expM\}$, with
\begin{equation}
\hat H^\expM=g\hat O_\Gamma\otimes \hat O_\Xi~,
\label{e.H_measure-like}
\end{equation}
where $\hat O_\Gamma=\sum_\pi\omega_\pi\proj{\pi}$ is the 
measured observable, while $\hat O_\Xi$ is the operator on ${\cal H}_\Xi$ 
conjugate to the pointer observable~\cite{Schlosshauer07}.
The resulting, measure-like, dynamics is such that decoherence of 
$\rho_\Gamma(t)$ w.r.t. the basis $\{\ket{\pi}\}$ implies 
$\rho_\Xi(t)=\sum_\pi|a_\pi|^2\proj{\Xi^\pi_t}$ with 
$\exval{\Xi^\pi_t|\Xi^{\pi'}_t}=\delta_{\pi\pi'}$ and $a_\pi$ such that
$\ket{\Gamma(0)}=\sum_\pi a_\pi\ket{\pi}$, and viceversa. Here $t$ 
indicates any time prior the output production when decoherence has 
already occurred.
This dynamics defines a CPTP map ${\cal E}^\expM$ via 
\begin{equation}
\proj{\Gamma}=\sum_{\pi\pi'}a_\pi a^*_{\pi'}\ket{\pi}\!\!\bra{\pi'}
\underset{{\cal E}^\expM}{\longrightarrow} 
\rho_\Xi=\sum_\pi|a_\pi|^2\proj{\Xi^\pi}~,
\label{e.measure-like_map}
\end{equation}
referred to as measure-and-prepare (m\&p) map in the literature. 
Notice that what characterizes ${\cal E}^\expM$ as a m\&p map is 
not the diagonal form of the output state $\rho_\Xi$, but rather the 
fact that its eigenvalues are constant and exclusively depend on the 
input state $\proj{\Gamma}$. 

\section{Large-$N$ as classical limit}
\label{a.large_N}
In order to define the classical limit of a quantum theory ${\cal Q}$ it 
is first necessary to identify a parameter $k$, usually dubbed 
"quanticity parameter", such that ${\cal Q}$ transforms into a classical theory 
${\cal C}$ as $k$ vanishes. By "transform" it is meant that a formal 
relation is set between Hilbert and phase spaces, Lie and Poisson 
brackets, Hamiltonian operators and functions. 
Consequently, the large-$N$ limit of ${\cal Q}$ implies a classical 
behaviour of the macroscopic system it describes IF $N\to\infty$ 
implies $k\to 0$. On the other hand, in order for this being the case 
it proves sufficient that GCS $\{\ket{\omega}\}$for ${\cal Q}$ exist and
feature some specific properties~\cite{Berezin1978,Yaffe82}. 
Amongst these, particularly relevant to this work is that
\begin{equation}
\lim_{k\to 0}
\,k\Big[\ln|\exval{\omega'|\omega}|\Big]\le 0~,
\label{e.Yaffe-assumption_3}
\end{equation}
where the equality holds iff $\omega=\omega'$, and the property implies 
the limit exists.
From the above property it follows\footnote{We use the Dirac-$\delta$ 
representation 
$\delta(x-y)=\lim_{\epsilon\to 0}(1/\epsilon)\exp\{(x-y)^2/\epsilon\}$.}
\begin{equation}
\lim_{k\to 0}
\frac{1}{k}|\exval{\omega|\omega'}|^2=\delta(\omega-\omega')~,
\label{e.GCS_orthogonal}
\end{equation}
which is a most relevant properties of GCS, namely that they become 
orthogonal in the classical limit.

It is worth mentioning that if ${\cal Q}$ features a global symmetry 
(also dubbed "supersymmetry" in the literature), GCS can be explicitly 
constructed and shown to feature the properties ensuring that the 
large-$N$ limit is indeed a classical one~\cite{Yaffe82}. However, 
whether the existence of one such symmetry be a necessary condition for 
a system to behave classically in the large-$N$ limit is not proven, 
although all of the known physical theories, be they vector-, matrix-, 
or gauge-theories, confirm the statement (see Sec.VII of 
Ref.~\cite{Yaffe82} for a thorough discussion about this point).
Incidentally, we believe the above supersymmetry be essential in 
defining what a macroscopic observer should actually be in order 
for Quantum Darwinism to occur, in a way similar to that discussed in 
Ref.~\cite{ChiribellaDA06} in the specific case of a quantum theory for
$N$ distinguishable particles with permutation global symmetry.

\section{Overlap between GCS and elements of an orthonormal basis in 
the large-$N$ limit} 
\label{a.large-N_Overlap}

One of the output of the GCS construction, and key-ingredient for their 
use, is the invariant measure $d\mu(\hat\omega)$ entering the identity 
resolution Eq.\eqref{e.ECS_resolid}. It is demonstrated~\cite{Yaffe82} 
that in order for such resolution to keep holding for whatever value of the 
quanticity parameter $k$ it must be $d\mu(\omega)=c_k dm(\omega)$, 
with $c_k$ a constant on ${\cal M}$ that depends on the normalization 
of the group-measure $d\mu(\hat\omega)$ and should be computed 
on a case-by-case basis. However, normalization of GCS is guaranteed by 
construction, and hence, via Eq.\eqref{e.ECS_resolid},   
\begin{equation}
\exval{\omega|\omega}=\int_{\cal M} 
c_k dm(\omega')|\exval{\omega|\omega'}|^2=1~,~\forall\ket{\omega}~;
\label{e.ECS_norm}
\end{equation}
Furthermore, from Eq.\eqref{e.GCS_orthogonal} it follows
$|\exval{\omega|\omega'}|^2\to k\delta(\omega-\omega')$ as $k$ vanishes, 
and hence
\begin{equation}
\lim_{k\to 0}c_k k\int_{\cal M}dm(\omega')\delta(\omega-\omega')=1~,
\label{e.ECS_norm_kto0}
\end{equation}
which implies $c_k=\frac{1}{k}$, 
as readily verified in those cases where an explicit form of GCS is 
available.
The fact that $c_k$ 
is independent of $\omega$ and goes like $\frac{1}{k}$ for vanishing $k$, 
enforces
\begin{equation}
\lim_{k\to 0} \int_{\cal M} \frac{1}{k}dm(\omega)
\exval{\xi^{'}|\omega}\exval{\omega|\xi{''}}
=\delta_{\xi^{'}\xi{''}}
\label{e.orthonormality_limit}
\end{equation}
to hold for whatever pair $(\ket{\xi^{'}},\ket{\xi{''}})$ of orthonormal 
states: as neither $dm(\omega)$ nor ${\cal M}$ depend on $k$, this is 
only possible if the two overlaps entering the integral are never 
simultaneously finite on ${\cal M}$ or, more precisely, on a set of 
finite measure. In other terms, Eq.\eqref{e.orthonormality_limit} 
implies Eq.\eqref{e.supports-dont-share}, and viceversa (which is 
trivial).

\end{document}